\documentclass[aps,prd,12pt,preprint,nofootinbib,superscriptaddress,titlepage,tightenlines,floatfix]{revtex4}
 
\usepackage{graphicx}
\usepackage{rotating}
\usepackage{dcolumn}
\usepackage{bm}
\usepackage{longtable}

\newcommand{\dilep}{\ell^+\ell^-}

\newcommand{\sprime}{s^{\,\prime}}

\begin{document}

\preprint{CLNS 05-1926}       
\preprint{CLEO 05-14}         

\title{\Large 
Observation of $\psi(3770)\to \pi\pi J/\psi$\\
and Measurement of $\Gamma_{ee}[\psi(2S)]$}

\author{N.~E.~Adam}
\author{J.~P.~Alexander}
\author{K.~Berkelman}
\author{D.~G.~Cassel}
\author{V.~Crede}
\author{J.~E.~Duboscq}
\author{K.~M.~Ecklund}
\author{R.~Ehrlich}
\author{L.~Fields}
\author{R.~S.~Galik}
\author{L.~Gibbons}
\author{B.~Gittelman}
\author{R.~Gray}
\author{S.~W.~Gray}
\author{D.~L.~Hartill}
\author{B.~K.~Heltsley}
\author{D.~Hertz}
\author{C.~D.~Jones}
\author{J.~Kandaswamy}
\author{D.~L.~Kreinick}
\author{V.~E.~Kuznetsov}
\author{H.~Mahlke-Kr\"uger}
\author{T.~O.~Meyer}
\author{P.~U.~E.~Onyisi}
\author{J.~R.~Patterson}
\author{D.~Peterson}
\author{E.~A.~Phillips}
\author{J.~Pivarski}
\author{D.~Riley}
\author{A.~Ryd}
\author{A.~J.~Sadoff}
\author{H.~Schwarthoff}
\author{X.~Shi}
\author{M.~R.~Shepherd}
\author{S.~Stroiney}
\author{W.~M.~Sun}
\author{D.~Urner}
\author{T.~Wilksen}
\author{K.~M.~Weaver}
\author{M.~Weinberger}
\affiliation{Cornell University, Ithaca, New York 14853}
% ...
\collaboration{CLEO Collaboration} 
\noaffiliation

\date{August 9, 2005}

\begin{abstract}
We observe signals for the decays $\psi(3770)\to X J/\psi$ 
from data acquired with the CLEO detector operating at 
the CESR $e^+e^-$ collider with $\sqrt{s}$=3773~MeV.
We measure the following branching fractions 
${\cal B}(\psi(3770)\to XJ/\psi)$ and significances:
$(189\pm20\pm20)\times 10^{-5}$ (11.6$\sigma$) for $X=\pi^+\pi^-$,
$(80\pm25\pm16)\times 10^{-5}$ (3.4$\sigma$) for $X=\pi^0\pi^0$, and
$(87\pm33\pm22)\times 10^{-5}$ (3.5$\sigma$) for $X=\eta$,
where the errors are statistical and systematic, respectively.
The radiative return process $e^+e^-\to\gamma\psi(2S)$ 
populates the same event sample and is used 
to measure $\Gamma_{ee}[\psi(2S)]=(2.54\pm0.03\pm0.11)$~keV.
\end{abstract}

\maketitle

\label{sec:intro}

The $\psi(3770)$ charmonium state decays most copiously into $D\bar{D}$~pairs,
but other decays similar to those of $\psi(2S)$ are 
predicted~\cite{quarkoniumtheory,yan,kuangyan3,lipkin,lane,kuangyan2,kuangyan1,rosnermixing}.
The $\psi(2S)$ mass eigenstate is expected~\cite{rosnermixing} have a
dominant $2\,^3S_1$ angular momentum eigenstate with a
small $1^3D_1$ admixture, and vice versa for the $\psi(3770)$.
Because more than half of $\psi(2S)$ decays contain a $J/\psi$ in the
final state, a $2\,^3S_1$ component enhances
similar transitions for $\psi(3770)$.
Theoretical estimates~\cite{kuangyan3,lane,kuangyan2,kuangyan1} of the
rate for transitions from the $1\,^3D_1$ eigenstate,
based on a QCD multipole expansion, span a broad range.
BES reported the first sighting of a $\psi(3770)$ {non-$D\bar{D}$}
decay~\cite{bespipijpsi},
at $\sim3\sigma$ significance,
with ${\cal B}(\psi(3770) \to \pi^+\pi^- J/\psi) = (0.34 \pm 0.14 \pm 0.09)\%$.

  In this Letter we describe a search for the $X J/\psi$ final states,
where $X=\pi^+\pi^-$, $\pi^0\pi^0$, $\eta$, and $\pi^0$, 
in $e^+e^-$ collision data taken at a center-of-mass energy
$\sqrt{s}$=3.773~GeV. We use $J/\psi\to\ell^+\ell^-$,
where $\ell^\pm\equiv e^\pm$ or $\mu^\pm$.
The data were acquired with the CLEO detector~\cite{cleoiiidetector} operating
at  
the Cornell Electron Storage Ring~\cite{cesr},
and correspond
to an integrated luminosity~\cite{LUMINS,babayaga} of ${\cal L}=(280.7\pm2.8)$~pb$^{-1}$.
The process $e^+e^-\to\gamma\psi(2S)$ dominates
this event sample and is treated as background;
it also yields a measurement of $\Gamma_{ee}[\psi(2S)]$.

  The primary background for $\psi(3770)\to X J/\psi$ is
the tail of the $\psi(2S)$ and radiative returns 
to it via initial state radiation (ISR)
({\sl i.e.} $e^+e^-\to\gamma\psi(2S)\to \gamma X J/\psi$);
the total radiated energy peaks
near $87$~MeV but can take on a range of values.
Similarly, there are radiative returns to that portion of
the $\psi(3770)$ lineshape lying below $\sqrt{s}$ which constitute
part of the signal. The differential
cross section for ${e^+e^-\to \gamma R}\to \gamma XJ/\psi$,
where $R=\psi(2S)$ or $\psi(3770)$,
can be expressed~\cite{KF,radcor,vectorisr} in terms of 
$XJ/\psi$ mass-squared $\sprime$ and
the scaled radiated energy 
$x\equiv 1-\sprime/s$ as
\begin{equation}
{{d\sigma}\over{dx}}= W(s,x)\times b(\sprime)\times F_X(\sprime)\times 
\Gamma_{ee}\times{\cal B}_X\ \ ,
\end{equation}
\noindent where $W(s,x)$ is the ISR $\gamma$-emission probability, 
$b(\sprime)$ is the relativistic Breit-Wigner formula, 
$F_X(\sprime)$ is the phase space factor~\cite{phasespace} appropriate
for $X$, $\Gamma_{ee}$ is the 
$e^+e^-$ partial width of $R$ (including vacuum polarization effects), 
and ${\cal B}_X\equiv{\cal B}(R\to XJ/\psi)$ 
signifies an exclusive branching fraction. The ISR kernel is,
at lowest order, 
\begin{equation}
W(s,x>x_0)\equiv {{2\alpha}\over{\pi
x}}\left(\ln{s\over{m_e^2}}-1\right)\left(1-x+{{x^2}\over2}\right)\, ,
\end{equation}
\noindent in which $x_0>0$ is a cutoff to prevent
the divergence of $\int W\, dx$, 
$m_e$ is the electron mass, and $\alpha$ is the fine
structure constant. The Breit-Wigner function is
\begin{equation}
b(s)\equiv { {12\pi \Gamma_R }\over{
(s-M_R^2)^2+ M_R^2\Gamma_R^2 } }\ ,
\end{equation}
\noindent in which $\Gamma_R$ is the full width
and $M_R$ the nominal mass. The phase
space factor is
$F_X(\sprime)\equiv {\left({p_X}/ p_0\right)}^{2L+1}$,
in which $p_X$ is the momentum of $X$ in the $R$ 
center-of-mass frame, $p_0$ is the value of $p_X$ at
$\sqrt{\sprime}=M_R$, and $L$ is the relative orbital
angular momentum between $X$ and $J/\psi$.
Eq.~(1) has one enhancement near $x=0$
due to the $1/x$ factor in $W(s,x)$, and, for $s$ sufficiently larger
than $M_R^2$, 
a much larger one near $x=1-M_R^2/s$, corresponding to
the peak of the Breit-Wigner resonance function.

  The cross section $\sigma(s)$ for
$e^+e^-\to \gamma R\to \gamma X J/\psi$
can be both obtained from Eq.~(1) and measured:
\begin{equation}
\sigma(s) = {{ N}\over{\epsilon\times {\cal L}}} =
\Gamma_{ee}\times{\cal B}_X\times I(s) \ ,
\end{equation}
\noindent in which $N$ is the number of events counted and $\epsilon$
is the detection efficiency obtained from Monte Carlo (MC)
simulation, and the integral
\begin{equation}
I(s)\equiv \int W(s,x)
 {{b(\sprime)}} F_X(\sprime)\, dx
\end{equation}
\noindent  is insensitive to the value of $\Gamma$. 
Hence a measurement of $\sigma(s)$ for $R=\psi(2S)$
can be combined with 
${\cal B}_X$ measurements~\cite{xjpsiprl} to yield 
$\Gamma_{ee}[\psi(2S)]$.

  We choose the $E_\gamma$ cutoff to be 
2~MeV ($x_0=1.06\times10^{-3}$),
small enough that events with $x<x_0$ 
are experimentally indistinguishable from
those with $x=x_0$. 
The expression~\cite{KF,radcor,vectorisr}  for $W(s,x<x_0)$ includes terms 
accounting for soft and virtual photon emission
(but not vacuum polarization, which is included
in $\Gamma_{ee}$); we obtain 
$I(s,x<x_0)=0.62\,b(s)\,F_X(s)$,
a result reproduced by 
the Babayaga~\cite{babayaga} $\mu^+\mu^-$ event generator.
The integral $I(s,x>x_0)$ can be performed numerically.
For $W(s,x>x_0)$, we employ the full expression including
higher order radiative corrections as given in Eq.~(28) of
Ref.~\cite{KF}; it gives values $\sim$19\% smaller than Eq.~(2)
for $J/\psi$ radiative returns from $\sqrt{s}=3.773$~GeV.
 
The {\tt EvtGen} event generator~\cite{EVTGEN}, which includes
final state radiation~\cite{PHOTOS}, and a 
GEANT-based~\cite{GEANT} detector simulation
are used to model the physics processes.
The generator implements a relative $S$-wave ($P$-wave) configuration 
between the $\pi\pi$ ($\eta$ or $\pi^0$) and
the $J/\psi$.
Radiative returns to $\psi(2S)$ and $\psi(3770)$ for $x>x_0$ are generated
with the polar angle distribution from Refs.~\cite{vectorisr}, 
and account for ISR according to Eqs.~(1-3). 
Separately, $X J/\psi$ events are also generated without
a photon to represent all $x<x_0$ events from $\psi(2S)$
and $\psi(3770)$, and are weighted with respect to the $x>x_0$
events according to the $I(s,x<x_0)/I(s,x>x_0)$ ratios.

Event selection implements the same requirements as
in the CLEO $\psi(2S)\to X J/\psi$
analysis~\cite{xjpsiprl} except for the changes described here. 
No $X$-recoil mass cuts are imposed. 
To increase acceptance for $J/\psi\to\ell^+\ell^-$, lepton candidates
at small polar angles ($0.85<|\cos\theta_\ell|<0.93$) are added.
A number of measures are taken to reduce backgrounds. We demand 
$m(\dilep) = 3.05 - 3.14$~GeV, and add to each lepton momentum vector 
any photon candidates located within a 100 mrad cone
of the initial lepton direction.
All $\pi^0$ candidates
must satisfy $m(\gamma\gamma)=110$-150~MeV.
For $X=\pi^+\pi^-$,
 neither pion candidate can be 
identified as an electron if $m(\pi^+\pi^-)<450$~MeV, which
suppresses $e^+e^-\to\ell^+\ell^-\gamma$, $\gamma\to e^+e^-$ events
in which the $e^+e^-$ pair from the photon conversion
is mistaken for the $\pi^+\pi^-$.
For $X=\pi^+\pi^-$ and $\pi^0\pi^0$,
we require $m(\pi\pi)>350$~MeV.
For the $\pi^0 J/\psi (\to e^+e^-)$ and 
$\eta (\to\gamma\gamma) J/\psi (\to e^+e^-)$ modes, 
background from Bhabha events 
is diminished by requiring $\cos\theta_{e^+}<0.3$.
For $X=\pi^0$ or $\eta (\to\gamma\gamma)$, 
radiative transitions from $\psi(3770)$ or the $\psi(2S)$ tail
to $\chi_{cJ}$ are suppressed by
requiring the least energetic photon in the $\pi^0$
or $\eta$ candidate to satisfy $E_\gamma >280$~MeV or $E_\gamma=30$-170~MeV.

  To extract the number of $\psi(2S)$ and $\psi(3770)$ events,
we fit the distribution of event missing momentum,
which can be interpreted as a measure of $E_\gamma$,
\begin{equation}
 k = {{s - M_J^2 + m_X^2 - 2\sqrt{s\,(p_X^2+m_X^2)}}
        \over{2\left(\sqrt{p_J^2 + M_J^2}-p_J\cos\phi \right)}}\,\, ,
\end{equation}
\noindent in which $M_J$ is the nominal $J/\psi$ mass, $p_J$ is the
measured dilepton momentum, $p_X$ is the measured $X$ momentum,
$m_X$ is the mass of $X$ (the PDG value~\cite{PDG} for $X=\eta\, , \pi^0$,
or the measured mass for $X=\pi\pi$), and $\phi$ is the measured
angle between the $J/\psi$ and the event missing momentum three-vector ({\bf k}).
The small ($\sim$2~mrad) crossing angle of
the incoming $e^\pm$ beams has been neglected.

 The phase space~\cite{phasespace} factor for $F_{\pi\pi}(s\,^\prime)$ 
is not as simple as
the $\left({p_X}/ p_0\right)^3$ dependence for $\eta$ and $\pi^0$
because the $\pi\pi$ mass varies. 
The average momentum of the $\pi\pi$ system
increases by $\sim$11\% 
from $\sqrt{\sprime}$=3.686~GeV to 3.773~GeV. 
As the $\pi\pi$ and $J/\psi$ are in a 
relative $S$-wave,
we set $F_{\pi\pi}$$(\sprime$=$(3.773~{\rm GeV})^2)$=$1.11$;
for other $\sprime$, $F_{\pi\pi}$  is scaled
linearly with $x$. The functional form of  $F_{\pi\pi}$ 
is not crucial because 
$d\sigma/dx$ is
small over the central portion of the interval $E_\gamma=0$-87~MeV.

  The distribution in $k$ for each exclusive
mode is subjected to a maximum likelihood fit for 
three components with floating normalizations:
a radiative return to $\psi(2S)$ shape obtained from MC simulation, 
a direct decay $\psi(3770)\to X J/\psi$ signal shape from MC simulation
(including radiative returns to the $\psi(3770)$ tail),
and a background component linear in $k$. 
Direct decays from
the $\psi(3770)$ and the tail of the $\psi(2S)$
add incoherently~\cite{kuangyan2}, so that the $\psi(2S)$ background
can be included without regard for interference.

\begin{figure}[thp]
\includegraphics*[width=6.5in]{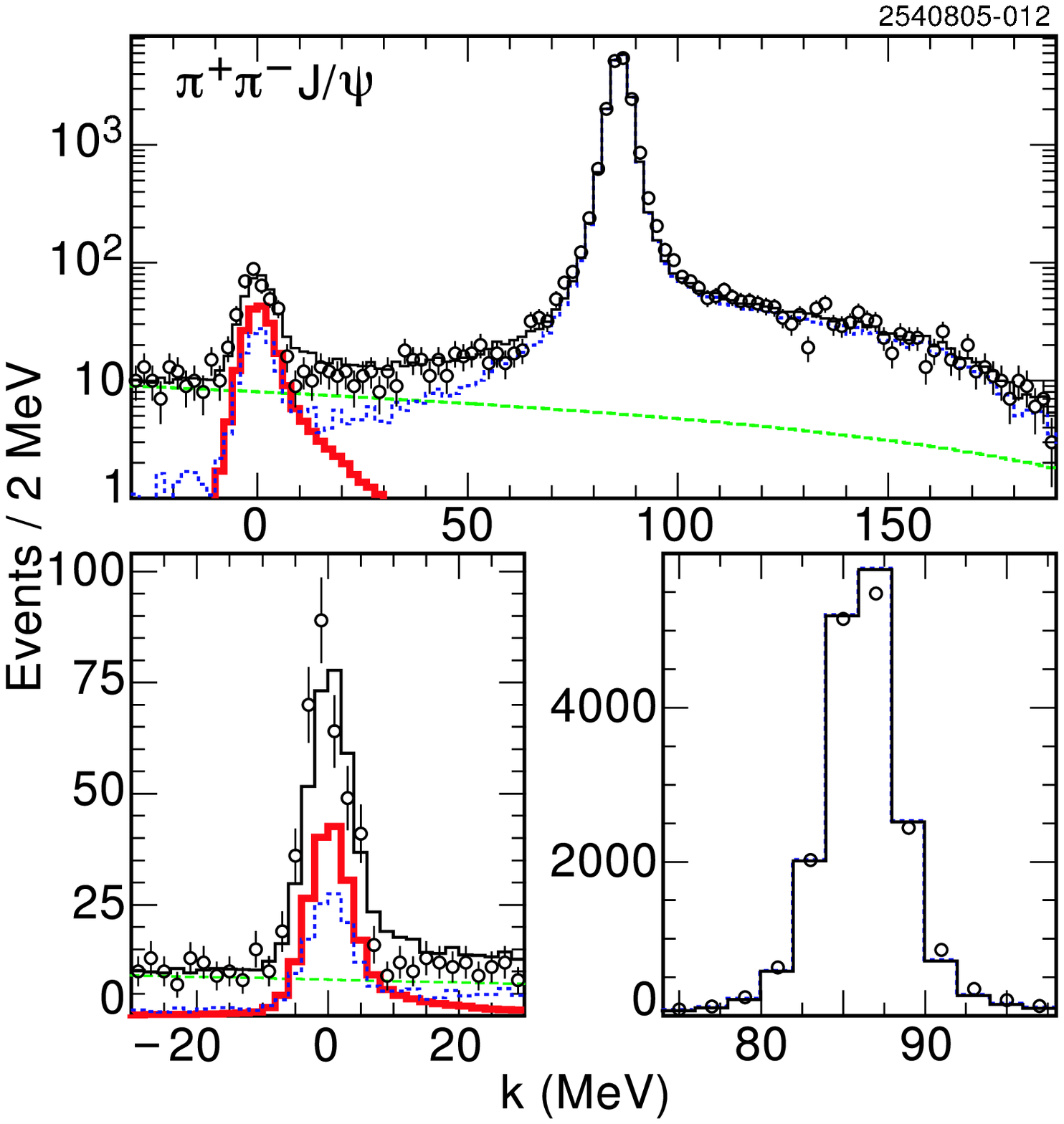}
\caption{Fit of the distribution in $k$
for the final state $\pi^+\pi^- J/\psi$, 
showing the data (open circles), overall fit (thin solid line),
direct $\psi(3770)$  decay peak (thick solid line),
radiative return to the $\psi(2S)$ (dotted line),
and the background term (dashed line), on
a logarithmic vertical scale (top), and on linear vertical scales focussed on
the direct decay peak (bottom left) and radiative return peak
(bottom right).\label{fig:pipitot} }
\end{figure}

\begin{figure}[tp]
\includegraphics*[width=6.5in]{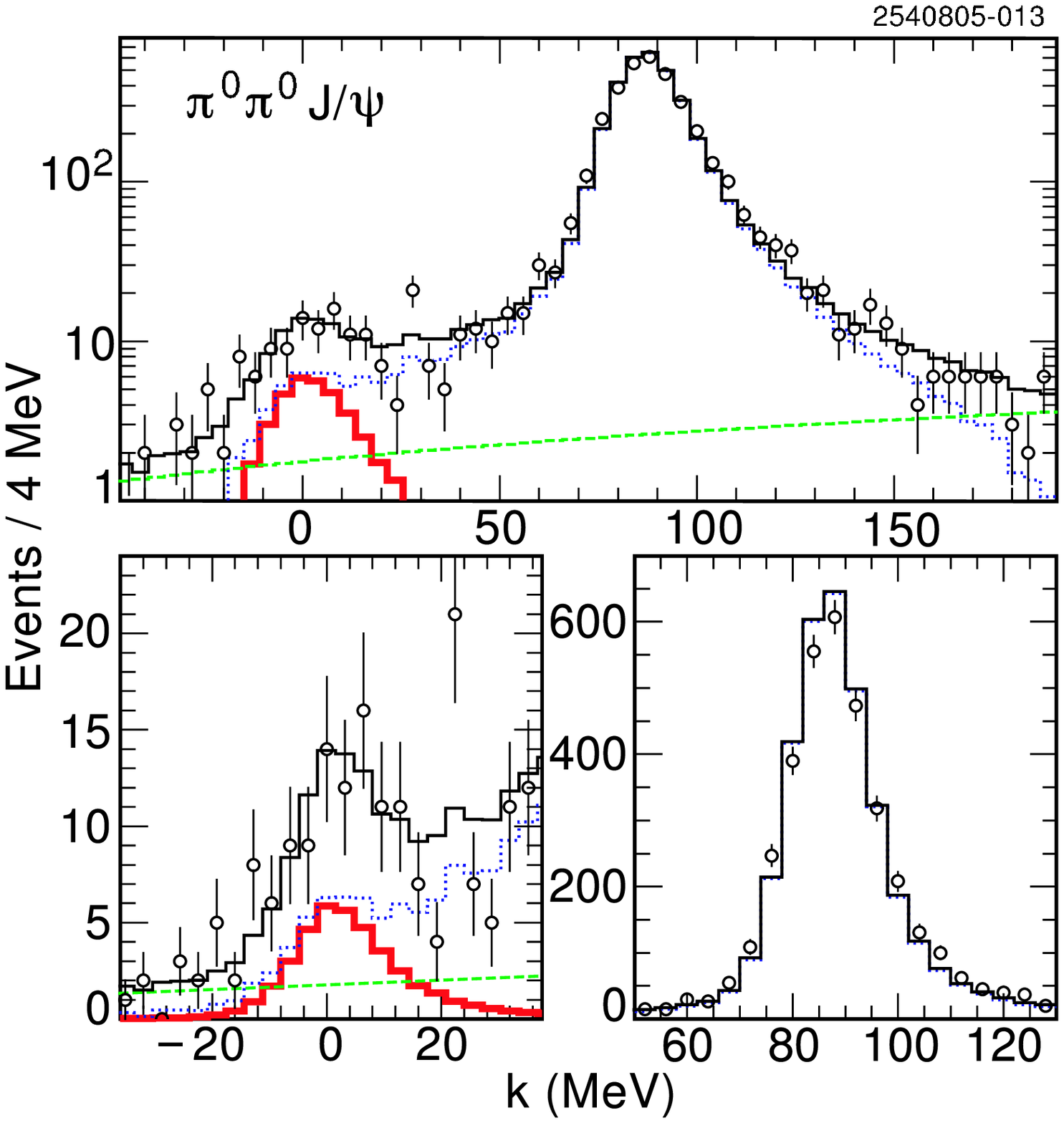}
\caption{Fit of the distribution in $k$
for the final state $\pi^0\pi^0 J/\psi$,
showing the data (open circles), overall fit (thin solid line),
direct $\psi(3770)$  decay peak (thick solid line),
radiative return to the $\psi(2S)$ (dotted line),
and the background term (dashed line), on
a logarithmic vertical scale (top), and on linear vertical scales focussed on
the direct decay peak (bottom left) and radiative return peak
(bottom right).\label{fig:pi0pi0tot} }
\end{figure}

\begin{figure}[thp]
\includegraphics*[width=6.5in]{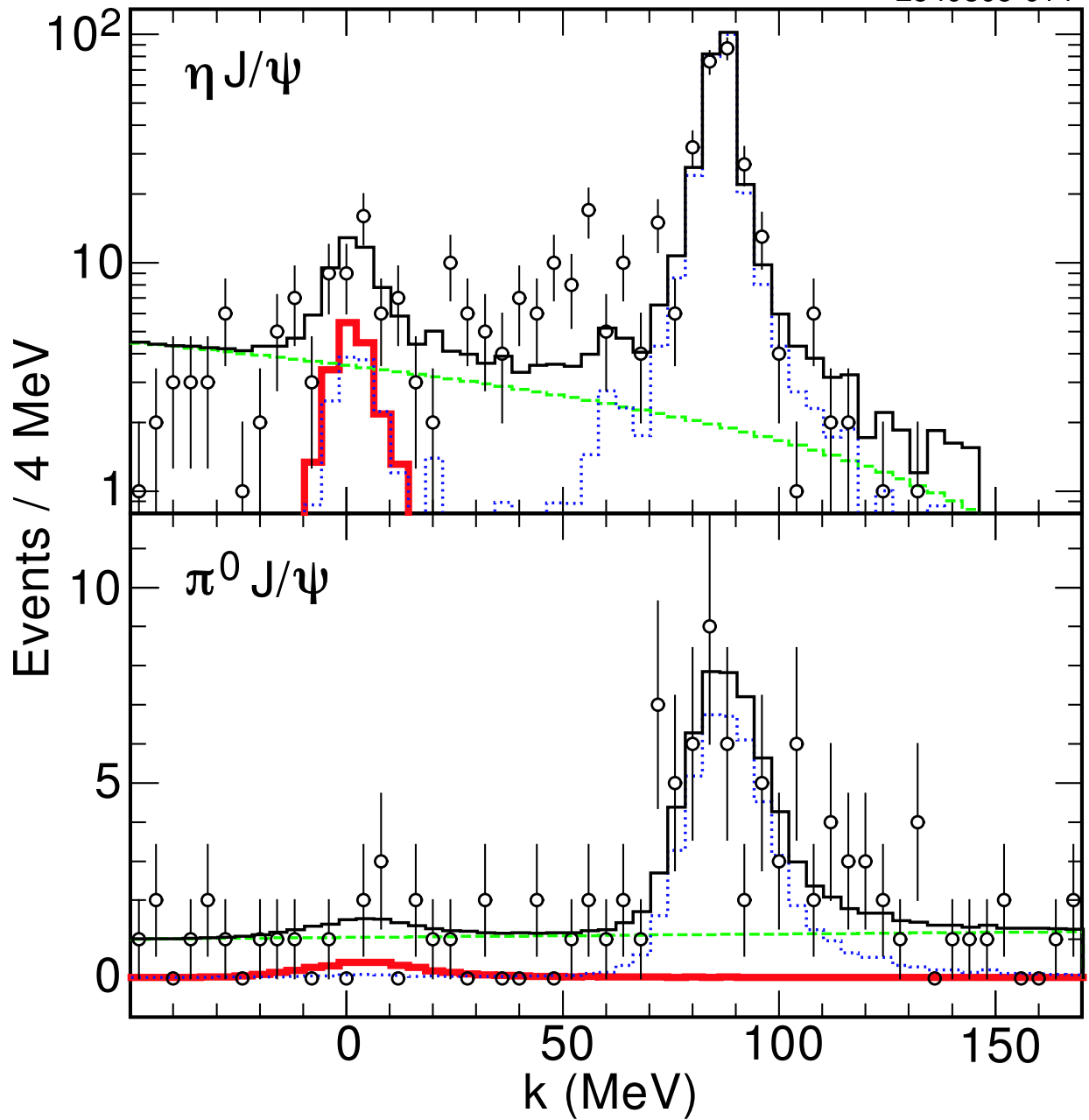}
\caption{Fit of the distribution in $k$
for the final state $\eta J/\psi$ (top) and $\pi^0 J/\psi$ (bottom),
showing for each 
the data (open circles), overall fit (thin line),
direct $\psi(3770)$  decay peak (thick solid line),
radiative return to the $\psi(2S)$ (dotted line),
and the background term (dashed line).\label{fig:etapi0tot} }
\end{figure}

  The distributions and fits are shown in 
Figs.~\ref{fig:pipitot}-\ref{fig:etapi0tot}. The fit
results and quantities derived from them appear in Table~\ref{tab:tableFits}.
The efficiencies include the MC correction factors 
from Ref.~\cite{xjpsiprl}, the visible cross sections
use ${\cal B}(J/\psi\to\ell^+\ell^-)$
from Ref.~\cite{dilepprd}, and the $\Gamma_{ee}$ values 
use the ${\cal B}(\psi(2S)\to XJ/\psi)$ results
from Ref.~\cite{xjpsiprl}. Statistical significances of
the $\psi(3770)$ signals, obtained from the differences
in log-likelihoods between fits with and without a signal component, 
are shown, indicating 
an unambiguous $\pi^+\pi^- J/\psi$ signal,
and suggestive $\pi^0\pi^0 J/\psi$ and $\eta J/\psi$ signals.
The product of the measured cross section $\sigma(D\bar{D})$~\cite{ddbarprl} 
and luminosity ${\cal L}$ is used to 
give the number of produced $\psi(3770)$ decays 
as $(1.80\pm0.03^{+0.04}_{-0.02})\times10^6$. 
Sidebands around $M(J/\psi)$ in the
dilepton mass distributions for events near the radiative return
peak in $k$ do not show evidence for additional background.
Feed-across background (from radiative returns to $\psi(2S)$ but with
non-signal $\psi(2S)$ decays) levels and uncertainties are determined 
using measured branching fractions~\cite{xjpsiprl} and MC simulation, as in
Ref.~\cite{xjpsiprl}.

The results for $\gamma\psi(2S)\to\gamma\pi^0 J/\psi$
are all treated as upper limits due to substantial
backgrounds from radiative Bhabha and muon pair events.
Efficiency-corrected, background-subtracted 
rates for $J/\psi\to e^+e^-$ and $J/\psi\to\mu^+\mu^-$
are consistent with each other. Allowing
second or third order polynomials in the background
parameterizations has a negligible effect upon $\Gamma_{ee}$.

Statistical errors dominate for the $\psi(3770)$ results
and systematic errors dominate for the $\psi(2S)$ results.
Table~\ref{tab:tableSys} summarizes the uncertainties 
that are uncorrelated for different $X$.
The systematic errors
on the fitted event yields include changes induced by variation of the
range in $k$ of the fit 
and the alternate use of a $\chi^2$ fit instead of
maximum likelihood.
The efficiency uncertainties are larger than 
in Ref.~\cite{xjpsiprl} 
because here the leptons are not restricted to $|\cos\theta_\ell|<0.83$,
the $\pi^0\to\gamma\gamma$ and $J/\psi\to\dilep$ mass cuts are tighter,
and we account for imperfections in the assumed
$\psi(2S)$ boost direction. 

Relative uncertainties that are correlated for all $X$
include those from ${\cal B}(J/\psi\to\ell^+\ell^-)$ (0.94\%
statistical, 0.71\% systematic), $I(s)$ ({\sl i.e.}  radiative
corrections) (2.0\%), ${\cal L}$ (1.0\%), and the
normalization portion of ${\cal B}_X$ (3.0\%).
The MC sample
used for this analysis has a mean and spread of $\sqrt{s}$
very close to that of the data, within 0.05~MeV and 0.02~MeV,
respectively, rendering negligible any remaining 
systematic effect upon $I(s)$. 

A single value, $\Gamma_{ee}[\psi(2S)]=(2.54\pm0.03\pm0.11)$~keV, 
is obtained by combining  
$\pi^+\pi^-J/\psi$, $\pi^0\pi^0J/\psi$, and $\eta J/\psi$ results, 
weighting each by the uncorrelated statistical and systematic errors.
The relative 4.4\% total uncertainty is dominated
by the common 3.0\% systematic normalization uncertainty 
in all CLEO $\psi(2S)$ branching fraction measurements~\cite{xjpsiprl}.
It is 2.5 standard deviations higher than and of comparable precision to
the PDG fit value, $(2.12\pm0.12)$~keV~\cite{PDG};
it is within two standard deviations of any
of the results obtained from 
scanning the $\psi(2S)$ peak, the most precise of which is the preliminary 
BES~\cite{besscan} result, $(2.25\pm0.11\pm0.02)$~keV.

\begin{figure}[thp]
\includegraphics*[width=6.5in]{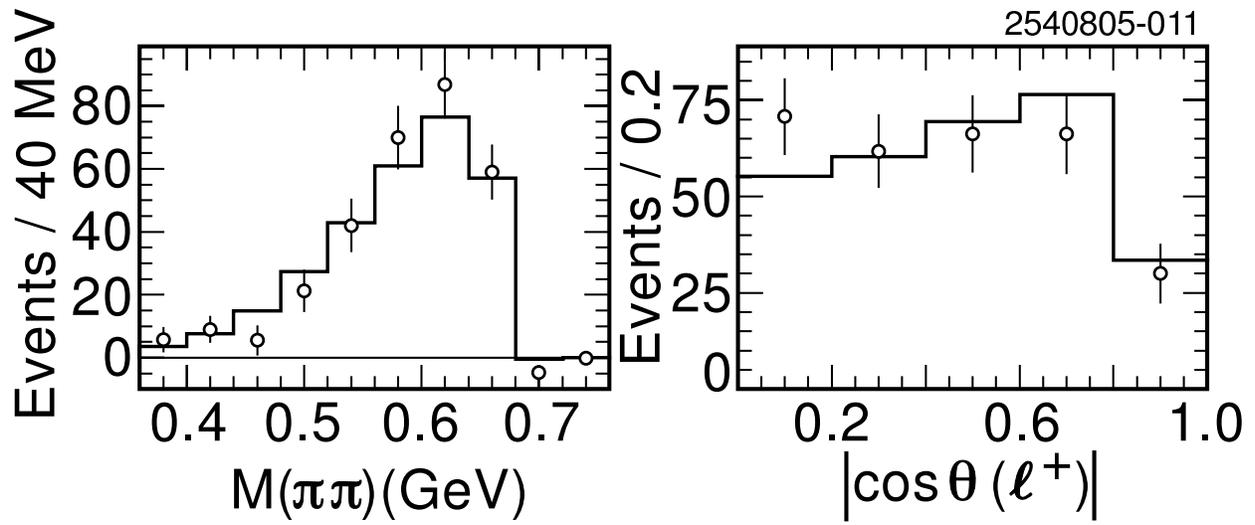}
\caption{Distributions in $\pi^+\pi^-\ell^+\ell^-$ events
of the $\pi^+\pi^-$ mass (left) and
polar angle (right) of the positively charged lepton from
data (open circles) and MC (solid line line). \label{fig:sigcth} }
\end{figure}

  Figure~\ref{fig:sigcth} shows the $m(\pi^+\pi^-)$ and 
$\ell^+$ polar angle distributions for 
$\pi^+\pi^- J/\psi$ events restricted to
$k=[-10,+10]$~MeV, background-subtracted 
with the sum of the $[-57,-17]$ and $[+13,+53]$~MeV sidebands
scaled down by a factor of four. The MC histograms
are normalized to the same areas as the data.
In both plots, neither data nor MC is corrected for detection efficiency.
The data points represent events from both
$\psi(3770)$ and $\psi(2S)$ in a ratio of $\sim$2:1.
The measured $m(\pi^+\pi^-)$ and $|\cos\theta (\ell^+)|$
distributions show consistency with 
the $\psi(2S)$-like $S$-wave MC predictions.

The branching
fraction for $\psi(3770)\to\pi^+\pi^- J/\psi$ is smaller than
that reported by BES~\cite{bespipijpsi}, 
but is consistent with it and more precise.
While the widths for $\psi(3770)\to\pi\pi J/\psi$
are in the broad range predicted by the QCD multipole
expansion models~\cite{yan,kuangyan1,kuangyan2,kuangyan3}, the $\pi\pi$
mass distribution appears to be much stiffer than
predicted for the large $D$-wave proportion featured in
these models. 
The branching fraction for  $\psi(3770)\to\pi\pi J/\psi$ also relates
to the interpretation of the $X(3872)$:
the small value does not strengthen the case
for conventional charmonium~\cite{bargodelq}.  The results combine to give
$\Sigma_X{\cal B}(\psi(3770)\to XJ/\psi)=(0.36\pm0.06)$\%, 
which corresponds to a cross section of $(23\pm5)$~pb. 
The 90\%~C.L. upper limits are  
${\cal B}(\psi(3770)\to\eta J/\psi ,\,\,\pi^0J/\psi)<0.15$\%, 0.028\%;
substantially more data would be required to quantitatively 
probe the $c\bar{c}$ purity of the $\psi(2S)$ and $\psi(3770)$
as proposed in Ref.~\cite{voloshin}.

In summary, we have observed the first statistically
compelling signal for non-$D\bar{D}$ decays of the  $\psi(3770)$, and with the
same data sample have achieved
improved precision on $\Gamma_{ee}[\psi(2S)]$.

We gratefully acknowledge the effort of the CESR staff 
in providing us with excellent luminosity and running conditions.
This work was supported by the National Science Foundation
and the U.S. Department of Energy.

\begin{turnpage}
\begin{table*}[thp]
\setlength{\tabcolsep}{0.4pc}
\catcode`?=\active \def?{\kern\digitwidth}
\caption{Results for radiative return process $e^+e^-\to\gamma\psi(2S)$, 
$\psi(2S)\to X J/\psi$ and direct decay $\psi(3770)\to XJ/\psi$. 
For each appears the fit yield $N$, efficiency $\epsilon$, 
and cross section $\sigma$.
In addition, for the radiative return process,
the integral $I(s)$ [followed by its value for $x<x_0$],
and the ${\cal B}(\psi(2S)\to X J/\psi )\times\Gamma_{ee}$ 
values inferred from $\sigma$ appear
along with the resulting $\Gamma_{ee}$.  
The bottom five rows include the significance in standard deviations 
of the $\psi(3770)\to XJ/\psi$ signals
and the $\psi(3770)$ branching fraction and partial width.
Errors shown are statistical and systematic, respectively.
}
\label{tab:tableFits}
\begin{center}
\begin{tabular}{c|cccc}
\hline
\hline
$X$ &
                  $\pi^+\pi^-$ & 
                  $\pi^0\pi^0$ & 
                  $\eta$       & 
                  $\pi^0$     \\\hline 
$N(\gamma\psi(2S)\to \gamma X J/\psi)$ &
                  $19469\pm145\pm195$ &
                  $3616\pm64\pm72$    &
                  $291\pm19\pm15$ &
                  $<37$ \\ 
$\epsilon (\gamma\psi(2S)\to \gamma X J/\psi)$ (\%)  &
                  $56.23\pm0.07\pm0.90$ &
                  $21.66\pm0.06\pm0.65$ &
                  $7.89\pm0.17\pm0.28$  &
                  $11.33\pm0.12\pm0.66$\\
$\sigma (\gamma\psi(2S)\to \gamma X J/\psi)$ (pb) &
                  $1036\pm13\pm23$ &
                  $500\pm10\pm19$  &
                  $111\pm8\pm8$ &
                  $<10$\\
$I(s)$ (pb/keV)&
                  1215.4~[6.7] &
                  1215.4~[6.7] &
                  1251.9~[34.2] &
                  1215.2~[8.9] \\
${\cal B}(\psi(2S)\to X J/\psi)\times\Gamma_{ee}$ (eV) &
                  $852\pm10\pm26$ &
                  $411\pm8\pm18$  &
                  $88\pm6\pm7$ & 
                  $<8$ \\
$\Gamma_{ee} [\psi(2S)]$ (eV) &
                  $2541\pm32\pm113$ &
                  $2488\pm54\pm138$ &
                  $2716\pm191\pm217$ & 
                  $<6.2\times10^3$ \\
$N(\psi(3770)\to X J/\psi)$ &
                  $231\pm24\pm23$ & 
                  $39\pm12\pm8$ &
                  $22\pm8\pm6$ & 
                  $<10$~@90\%~C.L.\\
Significance     & 11.6$\sigma$ & 3.4$\sigma$ & 3.5$\sigma$ & 0$\sigma$ \\
$\epsilon (\psi(3770)\to X J/\psi)$ (\%)  &
                  $57.05\pm0.16\pm0.91$ & 
                  $22.86\pm0.13\pm0.69$ &
                  $11.80\pm0.13\pm0.43$ & 
                  $16.02\pm0.15\pm0.93$ \\
$\sigma(\psi(3770)\to X J/\psi)$ (pb) &
                  $12.1\pm1.8\pm1.2$ &
                  $5.1\pm2.0\pm1.0$  &
                  $5.5\pm2.1\pm1.4$ & 
                  $<1.8$~@90\%~C.L.\\
${\cal B}(\psi(3770)\to X J/\psi)$ (10$^{-5}$) &
                  $189\pm20\pm20$ &
                  $80\pm25\pm16$  &
                  $87\pm33\pm22$ & 
                  $<28$~@90\%~C.L.\\
$\Gamma(\psi(3770)\to X J/\psi)$ (keV) &
                  $45\pm5\pm7$ &
                  $19\pm6\pm4$  &
                  $21\pm8\pm6$ & 
                  $<7$~@90\%~C.L.\\
\hline
\hline
\end{tabular}
\end{center}
\end{table*}
\end{turnpage}

\begin{table}[thp]
\setlength{\tabcolsep}{0.4pc}
\catcode`?=\active \def?{\kern\digitwidth}
\caption{Uncorrelated relative 
uncertainties in percent for the 
results in Table~I; for correlated errors, see text.
}
\label{tab:tableSys}
\begin{center}
\begin{tabular}{c|cccc}
\hline
\hline
$X$ & $\pi^+\pi^-$ & $\pi^0\pi^0$ & $\eta$ & $\pi^0$\\\hline
$\psi(2S)$  yield (stat)                   & 0.7 & 1.8 & 6.5 & 19 \\
$\psi(3770)$ yield (stat)                  & 10  & 31  & 36  & $-$ \\
$\psi(2S)$  yield (sys)                    & 1.0 & 2.0 & 5.0 & 10 \\
$\psi(3770)$ yield (sys)                   & 10  & 20  & 25  & $-$ \\
Efficiency                                 & 1.6 & 3.0 & 3.6 & 5.8\\
$\psi(2S)$ Feed-across                     & 0.1 & 0.1 & 3.0 & 50\\
${\cal B}(\psi(2S)\to XJ/\psi)$ stat       & 0.4 & 0.9 & 1.9 & 7.7\\
${\cal B}(\psi(2S)\to XJ/\psi)$ sys        & 1.3 & 1.8 & 1.6 & 7.1\\
\hline
\hline
\end{tabular}
\end{center}
\end{table}

\end{document}